\documentclass[journal]{IEEEtran}
\usepackage[T1]{fontenc}
\usepackage{cite}
\usepackage{amsthm,amsmath,amssymb,bm}
\usepackage{mathrsfs}
\usepackage{graphicx}
\usepackage{enumitem}
\usepackage{algorithm}  
\usepackage{algorithmic}  
\usepackage{array}
\usepackage{subfigure}
\usepackage{psfrag}
\usepackage{booktabs}
\usepackage{tabularx}
\usepackage{xcolor}
\usepackage[caption=false]{subfig}

\ifCLASSINFOpdf
\else
\fi


\begin{document}
\renewcommand{\qedsymbol}{}
%

\title{A Near Field Low Time Complexity Beam Training Scheme Based on Spatial Orthogonal Decomposition }

\author{
	Xiyuan~Liu,~\IEEEmembership{Graduate Student Member,~IEEE,}
	Qingqing~Wu,~\IEEEmembership{Senior Member,~IEEE,}
	Rui~Wang,~\IEEEmembership{Senior Member,~IEEE,}
	Jun~Wu,~\IEEEmembership{Senior Member,~IEEE,}
	\thanks{X. Liu is with the College of Electronics and Information Engineering, Tongji University, Shanghai 201804, China (e-mail: 1910670@tongji.edu.cn).
		Q. Wu is with the Department of Electronic Engineering, Shanghai Jiao Tong University, 200240, China (e-mail: qingqingwu@sjtu.edu.cn).
		R. Wang is with the College of Electronics and Information Engineering, Tongji University, Shanghai 201804, China, and also with the Shanghai Institute of Intelligent Science and Technology, Tongji University, Shanghai 201804, China (e-mail: ruiwang@tongji.edu.cn).
		J. Wu is with the School of Computer Science, Fudan University, Shanghai 200433, China, and also with the Shanghai Qi Zhi Institute, Shanghai 200030, China (e-mail: wujun@fudan.edu.cn).}	
}

\maketitle

%



\IEEEtitleabstractindextext{%
	\begin{abstract}
		
		With the application of high-frequency communication and extremely large MIMO (XL-MIMO), the near-field effect has become increasingly apparent. The near-field beam design now requires consideration not only of the angle of arrival (AoA) information but also the curvature of arrival (CoA) information. However, due to their mutual coupling, orthogonally decomposing the near-field space becomes challenging. In this paper, we propose a Joint Autocorrelation and Cross-correlation (JAC) scheme to address the coupling information between near-field CoA and AoA. First, we analyze the similarity between the near-field problem and the Doppler problem in digital signal processing, revealing that the autocorrelation function can effectively extract CoA information. Subsequently, utilizing the obtained CoA, we transform the near-field problem into a far-field form, enabling the direct application of beam training schemes designed for the far-field in the near-field scenario. Finally, we analyze the characteristics of the far and near-field signal subspaces from the perspective of matrix theory and discuss how the JAC algorithm handles them. Numerical results demonstrate that the JAC scheme outperforms traditional methods in the high signal-to-noise ratio (SNR) regime. Moreover, the time complexity of the JAC algorithm is $\mathcal O(N+1)$, significantly smaller than existing near-field beam training algorithms.
		
	\end{abstract}
	
	\begin{IEEEkeywords}
		Near-field, beam training, sigal subspace, channel parameter estimation, AoA, CoA, XL-MIMO.
\end{IEEEkeywords}}


\IEEEdisplaynontitleabstractindextext

%
\IEEEpeerreviewmaketitle

\section{Introduction}
Since Marzetta introduced the concept of MIMO \cite{DBLP:journals/bell/Marzetta15}, MIMO has evolved from Massive MIMO to Extremely Large MIMO (XL-MIMO) for two primary reasons. First, the frequency band of communication has expanded from the original sub-6 GHz to high-frequency bands like millimeter wave and terahertz. Due to the limited scattering and diffraction capabilities of high-frequency signals, XL-MIMO becomes crucial as it provides beamforming gain to compensate for path loss in high-frequency signal propagation. Simultaneously, the shorter wavelength of high-frequency signals allows the integration of larger MIMOs within a limited physical size. Second, the adoption of hybrid beamforming and intelligent reflecting surface technologies has significantly reduced the average cost of MIMO \cite{DBLP:journals/twc/ZhengZ23, DBLP:journals/twc/WuZ19}. Consequently, XL-MIMO is deemed necessary, feasible, and practical in the 6G communication scenario.

However, XL-MIMO can introduce significant near-field effects \cite{DBLP:conf/iccchina/PengZJLL23, DBLP:conf/acssc/TorresSB20}. It is widely acknowledged that the boundary between the far-field and near-field is defined by the Rayleigh distance, expressed as $\frac{2D^2}{\lambda}$, where $D$ represents the array size, and $\lambda$ is the wavelength of the signal. In XL-MIMO and high-frequency communication scenarios, the larger $D$ and smaller $\lambda$ result in a substantially increased Rayleigh distance compared to traditional communication scenarios \cite{DBLP:conf/wcsp/WangZZTY21,DBLP:conf/pimrc/HuIW22}. This shift causes many users initially considered in the far-field in low-frequency scenarios to be situated in the near-field in high-frequency scenarios. The near-field effect introduces significant challenges to beamforming design, codebook design, and the beam training process, emerging as the primary bottleneck in current high-frequency communication and XL-MIMO scenarios.

The main distinction between the near-field and far-field lies in the fact that electromagnetic waves exhibit spherical wave characteristics in the near-field, as opposed to plane waves in the far-field. Consequently, in the far-field, electromagnetic wave characteristics can be solely described through the angles of arrival (AoA) or departure (AoD). \footnote{This paper primarily focuses on the uplink channel; hence, only the case of an array receiving signals will be considered in the subsequent text.} In the near-field region, owing to the characteristics of spherical waves, the AoAs observed by antennas at different positions of the array vary, and the AoA demonstrates a regular function with positional changes \cite{10146329}. To elucidate the AoA variations, the introduction of the concept of curvature of arrival (CoA) becomes necessary to assist in delineating the spatial characteristics of electromagnetic waves in near-field channels \cite{DBLP:journals/tsp/Friedlander19a}. Generally, the CoA of spherical waves in the near-field region can be considered constant \cite{DBLP:journals/ojcs/BjornsonS20}. Therefore, combining CoA information with AoA information at the reference antenna enables the description of the characteristics of the near-field channel.

However, the introduction of CoA significantly complicates the near-field problem compared to the far-field problem. There are two primary challenges in beam training within near-field scenarios. First, the joint estimation of CoA and AoA introduces a multiplicative time complexity, resulting in a quadratic form of the number of antennas. This complexity arises because the array's resolution for both AoA and CoA is directly proportional to the ratio of the array's physical size to wavelength, denoted as $\frac{D}{\lambda}$. In the context of XL-MIMO, the time complexity of separately estimating AoA or CoA is $\mathcal{O}(N)$, where $N$ represents the number of antennas in the array. When estimating AoA and CoA simultaneously, the time complexity becomes $\mathcal{O}(N^2)$ due to their combination in forming electromagnetic waves of different shapes. Given that XL-MIMO involves a larger value of $N$ compared to Massive MIMO, the near-field beam training problem becomes exceptionally time-consuming, exacerbated by the quadratic time complexity and larger $N$. Consequently, designing an efficient beam training scheme in the near-field becomes a considerable challenge. The second primary challenge faced by near-field beam training is the coupling of CoA and AoA information, presenting difficulties in orthogonal decomposition of the near-field space. To the best of our knowledge, no orthogonal decomposition scheme in the near-field has been proposed thus far.

However, if orthogonal decomposition cannot be achieved in the near field, it can lead to the following two main problems. The first problem is the further increase in the time complexity of near-field beamforming. This arises because, without the ability to orthogonally decompose space, it becomes impossible to explore the entire space with a finite number of beams without duplication or leakage. Consequently, it becomes necessary to repeat a certain area to enhance the spatial coverage of the beam training process, introducing significant redundancy and consuming additional time. The second challenge arises from the invalidation of certain indicators used to measure beam training and codebook design. The current indicators, including average reachability rate and average mismatch rate, are designed to assess the algorithm's ability to resist random errors caused by noise. However, they are inadequate for measuring the impact of deterministic errors resulting from the algorithm's inability to cover all spaces. \footnote{In near-field problems, the lack of spatial decomposition schemes affects the algorithm's error rate, considering both noise and the spatial coverage of the algorithm. It's crucial to recognize the clear distinction between random and deterministic errors in research. For instance, addressing beam mismatches due to random errors can be mitigated by conducting additional attempts. On the contrary, if the codebook lacks a beam covering the user's area, retraining becomes futile. Therefore, the use of performance indicators with average meaning to describe algorithm performance in the presence of deterministic errors can interfere with the algorithm's practical use.}

To address these gaps, we integrate digital signal processing (DSP) and matrix theory to analyze the channel characteristics of the near-field channel and propose decoupling and CoA and AoA methods to achieve orthogonal decomposition in the near-field space. Additionally, we propose a new near-field beam training scheme using the above principles. Our main contributions are as follows:
\begin{itemize}
	\item
	Due to the mathematical similarity between DSP and array signal processing (ASP), we define the near-field problem as a spatial Doppler problem. Considering the direction of the beam as the magnitude of the spatial frequency, spherical waves in near-field problems can be viewed as a phenomenon of spatial frequency shift, akin to the Doppler frequency shift phenomenon. This modeling approach allows the application of numerous algorithms for Doppler frequency offset estimation in DSP to near-field CoA estimation problems. Simultaneously, we delve into the analysis of dual concept pairs in other ASP and DSP, laying the groundwork for future investigations into the spatial Doppler phenomenon.
	
	\item
	We propose a method to decouple CoA and AoA information. Based on that, we introduce a novel near-field beam training scheme, namely the Joint Autocorrelation and Cross-correlation (JAC) scheme. First, we observe that while the cross-correlation function between the channel and the codeword is related to both the CoA and AoA of the channel, the autocorrelation function of the channel is only associated with the CoA information and not with AoA. Therefore, we present a solution to determine CoA based on the autocorrelation function of the received signal, achieving the complete decoupling of CoA and AoA. Second, leveraging the estimated CoA information, we employ the cross-correlation method to estimate AoA. We demonstrate that the AoA estimation in this case is entirely equivalent to the far-field, enabling the use of far-field beam training methods in this stage. The entire two-step beam training scheme is termed the JAC scheme, boasting a time complexity of $\mathcal{O}(N+1)$, significantly lower than existing near-field beam training schemes.
	
	\item
	We analyze differences in signal subspaces between the far-field and near-field from a matrix theory perspective. We clarify the rationale behind the JAC algorithm based on this analysis. The examination from the matrix theory perspective also illuminates why the cross-correlation method faces challenges in achieving complete orthogonal decomposition in the near-field, while the JAC method excels in this regard.
	
	\item
	Finally, we compare the JAC method with other beam training schemes, and the numerical results indicate that our scheme achieved a higher average achievable rate under a high signal-noise ratio (SNR), demonstrating the JAC scheme's superior spatial coverage.

\end{itemize}

The remainder of this paper is organized as follows. Section II presents the system model. Section III introduces the JAC scheme, while Section IV analyzes the processing and advantages of the JAC scheme in matrix theory. Section V demonstrates the performance of the JAC algorithm through simulation experiments. The conclusion is presented in Section VI.

{\itshape Notations}: In this paper, scalars, vectors, and matrices are denoted by italic letters, bold-face lower-case, and upper-case letters, respectively. The space of ${ x} \times { y}$ complex-valued matrices is denoted by $\mathbb{C}^{{x} \times { y}}$. For a complex-value vector $\bm x$, ${\bm x}\otimes{\bm y}$ denotes the Kronecker product of $\bm x$ and $\bm y$ while $|{\bm x}|$ denotes its modulus and diag$({\bm x})$ denotes a diagonal matrix with each diagonal entry being the corresponding entry in $\bm x$. ${\bm x}\cdot{\bm y} $ denotes the dot product between these two vectors, while the cross product between $\bm x$ and $\bm y$ is represented by ${\bm x}{\bm y}$ or ${\bm x}\times {\bm y}$. $\vert {\bm x} \vert$ is the modulus of ${\bm x}$. For a function ${\bm y}=H({\bm x})$, $H^{-1}({\bm y})$ denotes its inverse function. For a general matrix $\bm A$, ${\bm A}^*$,${\bm A}^H$, and ${\bm A}[i,j]$ denote its conjugate, conjugate transpose, and the $(i,j)$th entry, respectively. $\jmath$ denotes the imaginary unit, i.e., ${\jmath}^2=-1$.



\section{System model }
\begin{figure}
	\centering
	\includegraphics[scale=1]{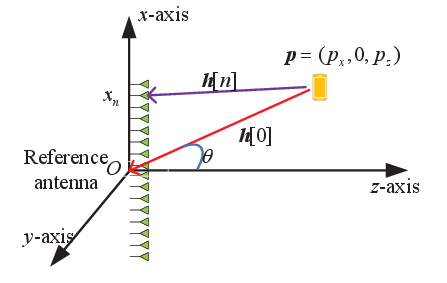}
	\caption{ Near-field channel model for ULA communication system.}     \label{coordinate system}
\end{figure}
As shown in Fig. \ref{coordinate system}, we first consider the uniform linear array (ULA) in MIMO scenarios. The base station is a ULA with $N$ antennas on the x-axis, with the user at point ${\bm p}=(p_x,0,p_z)$ in the $xOz$ plane. The origin point is located at the midpoint of the two antennas in the middle of the array.\footnote{Only the case where $N$ is an even number is considered here. However, since the parity of $N$ does not affect the conclusion of this paper, there is no separate discussion regarding the case where $N$ is odd or even.} The angle between the line connecting the user and the origin and the $x$-axis is $\theta$. $n$ is the index of the antenna, and ${\bm x}_n=(x_n,0,0)$ is the coordinate of the $n$th antenna. $x_n$ can be gotten as:
\begin{equation}
	x_n=\Bigg(n-\frac{N+1}{2}\Bigg)d,
\end{equation}
where $d=\frac{\lambda}{2}$ is the spacing distance of the array. $\lambda$ is the wavelength of the carrier. This paper considers uplink channels. The received signal can be represented as:
\begin{equation}\label{rn}
	\begin{split}
		{\bm r}[n] &={\bm h}[n]+n_{\sigma}\\
		& =e^{{\jmath}k\left(\vert {\bm p}-{\bm x}_n\vert-\vert {\bm p} \vert\right)}+n_{\sigma},
	\end{split}
\end{equation}
where $n_{\sigma}$ is the channel noise. $k=\frac{\omega}{c}=\frac{2\pi}{\lambda}$ where $\omega$, $c$, $\lambda$ are carrier frequency, light speed and wavelength of the carrier, respectively. We use the binomial theorem to expand $\vert {\bm p}-{\bm x}_n\vert$: \footnote{Expanding using Taylor's formula will result in the same form \cite{DBLP:journals/tcom/CuiD22}, which does not affect the conclusions later in this paper.}
\begin{equation}\label{binomial theorem}
	\begin{split}
		\vert {\bm p}-{\bm x}_n\vert&={ \vert {\bm p}\vert+\left(-2r \sin \theta x_n +x_n^2 \right)}^{\frac{1}{2}}\\
		&= \vert {\bm p} \vert-x_n \sin \theta + \frac{x_n^2 \cos^2\theta}{2\vert {\bm p}\vert}\\
		&+\frac{x_n^3 \cos^2 \theta \sin\theta}{2\vert{\bm p}\vert^2}+\cdots.
	\end{split}
\end{equation}

Based on the relationship with $x_n$, we categorize the terms in equation (\ref{binomial theorem}) as constant, linear, quadratic, and higher-order, respectively. Generally, we consider the phase error of ${\bm r}[n]$ negligible when it is less than $\frac{\pi}{8}$. The omitted portion of formula (\ref{binomial theorem}) represents the infinitesimal of the first few terms; hence, our discussion will focus on these initial terms.

When $\frac{x_n^3 \cos^2\theta \sin\theta}{2\vert{\bm p}\vert^2}>\frac{\lambda}{16}$, leading to $\vert {\bm p}\vert <0.62\sqrt{\frac{(Nd)^3}{\lambda}}$ \cite{7942128, DBLP:conf/acssc/BjornsonDS21, DBLP:journals/twc/JiangGJZZ23 }, electromagnetic waves exhibit electrical resistance \cite{DBLP:conf/pimrc/HuIW22} and cannot radiate energy outward \cite{DBLP:conf/pimrc/HuIW22}. Consequently, in communication scenarios, as long as the signal can propagate, it must satisfy $\frac{x_n^3 \cos^2\theta \sin\theta}{2\vert{\bm p}\vert^2}<\frac{\lambda}{16}$. Hence, higher-order terms can always be neglected in the communication channel model.

When $\frac{x_n^2 \cos^2\theta}{2\vert {\bm p}\vert}<\frac{\lambda}{16}$ holds, $\vert{\bm p}\vert>\frac{2(Nd)^2}{\lambda}$. This condition characterizes the Fraunhofer region, also known as the far-field region, where ${\bm h}[n]$ becomes a function of $\theta$ independent of $\vert{\bm p}\vert$. Conversely, when $\vert{\bm p}\vert<\frac{2(Nd)^2}{\lambda}$, the corresponding region is termed the Fresnel or near-field region. Thus, from a mathematical model perspective, considering the relationship with $x_n$, the phase function of the near-field channel is quadratic, while the far-field channel is a linear function. Subsequently, we will analyze the primary distinctions between the far and near-fields from the viewpoint of the mathematical duality between digital signal processing (DSP) and array signal processing (ASP) to formulate research approaches in the near-field.

In the far-field, the channel model is expressed as follows:
\begin{equation}\label{far-field channel}
	{\bm h}_{\rm far}[n]=e^{-{\jmath}kdx_n \sin\theta},
\end{equation}
which can be interpreted as a sampling function for a single-frequency signal \cite{DBLP:journals/tsp/PizzoTSM22}. Here, the spacing distance $d$ denotes the sampling interval in space, and the array size $D=Nd$ represents the sampling length. Analogous to the concept of sampling time length in DSP, we define $D$ as the length of the sampling space. The projection value of the wave number in the $x$-direction is $k\sin\theta$, representing the spatial frequency of the signal. Therefore, when $k$ is known, determining the AoA is akin to determining the spatial frequency of the signal. In the following discussion, we refer to the angle of the channel or beam as the spatial frequency.

In the far-field scenario, the spatial frequency of the signal is a fixed value. Therefore, the beamforming problem in the far-field essentially becomes a spectral analysis problem of stationary random signals. Furthermore, because the received signal is constrained by the array size in space, it is equivalent to multiplying a gate function by the infinitely received signal in space. Consequently, the spatial frequency spectrum of the signal is convolved with a sinc function. The width of the main lobe of this sinc function is the spatial coherence bandwidth, also known as the beam width.

In DSP, the coherent bandwidth of the channel depends on the root mean square delay expansion of the channel. In ASP, the beamwidth of the array also depends on the spatial expansion of the array, i.e., the size of the array. Due to the presence of spatial coherent bandwidth, continuous Fourier transforms on the signal are unnecessary. Instead, we only need to solve for the corresponding Fourier series within each coherent bandwidth. Generally, we assume there are $N$ independent Fourier series on an array of $N$ antennas, i.e., the coherent bandwidth of the array is $\frac{2}{N}$. The process of solving Fourier series involves convolving the received signal one by one with single-frequency signals. This single-frequency signal is known as the steering vector. Assuming the steering vector towards the angle $\theta_0$ is:
\begin{equation}
	{\bm a}_{\theta_0}[n]=e^{{\jmath}kx_n \sin\theta_0}.
\end{equation}
Then the coefficient of its corresponding Fourier series is:
\begin{equation}
	{\mathcal F}_{\theta_0}=\frac{1}{N}\Bigg| \sum_{n=1}^N {\bm a}_{\theta_0}[n] {\bm r}[n]\Bigg|.
\end{equation}
It is worth mentioning that because each Fourier series corresponds to an AoA, only the coefficients of $L$ Fourier series are significantly greater than 0, where $L$ is the number of multipaths. The other coefficients close to 0 are all generated by noise. So we can represent the channel in the form of $L$ series \cite{DBLP:journals/twc/PizzoSM22}:
\begin{equation}\label{compressed channel}
	{\bm h}_{\rm far}[n]=\sum_{l=1}^L \mathcal{F}_{\theta_l} \sum_{n=1}^N {\bm a}_{\theta_l}^*[n].
\end{equation}
Generally, compressed sensing algorithms can achieve the formula (\ref{compressed channel}). Although it is theoretically possible to achieve a series selection greater than 1 for analog beamforming, in reality, this method is not optimal because analog beamforming ultimately mixes all signals together. Therefore, from the perspective of practical applications, we can say that the number of Fourier series corresponding to analog beamforming is 1. Similarly, hybrid beamforming should select the largest $N_{\rm RF}$ Fourier series, where $N_{\rm RF}$ is the number of RF chains. Therefore, from the perspective of practical applications, the number of RF chains in ASP corresponds to the number of Fourier series in DSP.

\begin{table}[!htbp]\label{aspdsp}
	\centering
	\caption{\textsc{The Dual Concept between Digital Signal Processing and Array Signal Processing.}}
	\begin{tabularx}{\linewidth}{|p{4.2cm}|p{3.78cm}|}
		\hline
		\textbf{ Array signal processing } &  \textbf{Digital signal processing }  \\
		\hline
		Space domain & Time domain \\
		\hline
		Spacial domain & Frequency domain \\
		\hline
		Far-field signal & Stationary random signal\\
		\hline
		Near-field signal & Non-stationary random signal \\
		\hline
		Number of antennas & Number of sampling points \\
		\hline
		Spacing distance & Sampling frequency \\
		\hline
		Array size & Sampling  time  \\
		\hline
		Beam width & Coherence bandwidth\\
		\hline
		Array length that can be seen as far-field & Coherence time \\
		\hline
		MIMO & Multipath effect \\
		\hline
		The number of RF chains & The number of Fourier series \\
		\hline
		Beamforming & Frequency domain filtering \\
		\hline
		Optimal beamforming design & Signal recovery \\
		\hline
	\end{tabularx}%
	\label{t1}%
	
\end{table}%

Below, we incorporate the conclusions from the far-field into the near-field and analyze the differences in near-field channels from a signal processing perspective. Based on the above analysis, the channel model for the near-field is given by:
\begin{equation}\label{near_ori_channel}
	{\bm h}_{\rm near}[n]=e^{{\jmath}k\left(-x_n \sin\theta+\frac{x^2_n\cos^2\theta}{2\vert{\bm p}\vert}\right)}.
\end{equation}
To simplify the representation, we introduce the parameters $p_1=\frac{\cos^2\theta}{\vert{\bm p}\vert}$ and $p_2=-\sin\theta$. Using these two parameters, formula (\ref{near_ori_channel}) can be expressed as:
\begin{equation}\label{sim_near}
	{\bm h}_{\rm near}[n]=e^{{\jmath}k\left(\frac{p_1}{2}x^2_n+p_2 x_n\right)}.
\end{equation}
In light of previous far-field conclusions, the first-order derivative value on the channel phase represents the spatial frequency value of the channel. In the far-field, the spatial frequency is a fixed value on the array, whereas in the near-field, the spatial frequency is a linear function of $x_n$. This aligns with the physical interpretation of the near-field, indicating that the channel angles observed by antennas at different positions of the array vary. Consequently, the near-field problem is essentially a challenge involving spatially non-stationary random signals \cite{DBLP:journals/wc/CarvalhoAAAH20}. In this paper, we refer to the near-field effect as the spatial Doppler phenomenon. While the time-domain Doppler phenomenon describes the frequency of the signal changing with time, the spatial Doppler phenomenon characterizes the signal's angle changing with space. For clarity, we rewrite formula (\ref{sim_near}) in Doppler form:
\begin{equation}\label{doppler_near}
	{\bm h}_{\rm near}[n]=e^{{\jmath}k\left(p_2+\Delta p_2(p_1,x_n)\right)x_n}.
\end{equation}
From formula (\ref{doppler_near}), it is evident that the core of the near-field problem lies in the estimation of Doppler shift and spectrum for non-stationary random signals. To address this, we must first estimate the Doppler frequency shift's value, then employ this value to restore the spectrum of the signal itself, and finally conduct spectral analysis.
To elucidate the essence of signal processing in near-field problems, Table I compares various concepts of digital signal processing and array signal processing.

\section{Joint Autocorrelation and Cross Correlation Algorithm}
In this section, we introduce a novel near-field beamforming algorithm, termed the Joint Autocorrelation and Cross-correlation algorithm (JAC algorithm), based on the near-field principles outlined in Section II.

The challenge of beamforming and position estimation in the near-field fundamentally involves the estimation of ${\bm p}$. In conventional estimation algorithms, $\bm p$ is typically considered a function of the user's distance and angle relative to the reference antenna of the array \cite{DBLP:journals/wcl/ZhangWY22, DBLP:journals/icl/GanHYZZ23}. However, in this paper, we treat $\bm p$ as a function of two parameters, $p_1$ and $p_2$. Although these two representations are equivalent, the complexity of designing algorithms differs significantly. We can compare Digital Signal Processing (DSP) and Array Signal Processing (ASP) to derive estimation methods for $p_1$ and $p_2$. Given our emphasis on signal characteristics over noise, the subsequent discussion omits the noise term in Formula (\ref{rn}).

We have the following proposition based on the analysis of the duality between DSP and ASP in the previous section.

{\itshape Proposition 1 (Near Field First Equivalent Principle):} The process of solving $p_1$ is independent of $p_2$ according to the equivalent relationship between the near-field problem and the Doppler frequency shift problem. $p_1$ can be directly obtained by determining the size of the interference space.

{\itshape proof :} See Appendix \ref{p1proof}. $\hfill\blacksquare$

\begin{figure}
	\centering
	\includegraphics[scale=1]{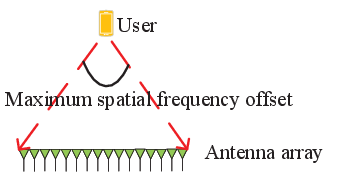}
	\caption{ Spatial Doppler phenomenon.}     \label{spacial_frenquency}
\end{figure}

We also term the first equivalent principle in the near-field as the decoupling principle. Proposition 1 asserts that the near-field problem in ASP is essentially analogous to the Doppler frequency shift problem in DSP. As depicted in Fig. \ref{spacial_frenquency}, the origin of the near-field problem lies in the distinct angles of arrival (AoA) of signals received by antennas at various positions in the array. Once the angle offset is known, the distance from the user to the array can be determined. The angle offset essentially translates to the difference in signal spatial frequency. Therefore, solving $p_1$ is tantamount to determining the maximum offset of signal spatial frequency. To address the spatial frequency offset of the signal, we need to solve for the coherent space of the signal. The autocorrelation method is more suited for resolving coherent space compared to the cross-correlation method.\footnote{Here, the autocorrelation method involves multiplying two time or spatial segments of a signal to ascertain the rate of change of signal frequency over time or space. Although this method isn't apt for solving signal frequency, it proves highly suitable for determining the rate of change of signal frequency. On the other hand, the cross-correlation method involves employing pre-set beamforming schemes and channels for cross-correlation, where the magnitude of the cross-correlation value directly increases the power of the received signal. A large cross-correlation value indicates that the channel aligns with the beamforming scheme. While many existing beam training methods fall under the cross-correlation method, not all variables are suitable for determining using this method.}
We define
\begin{equation}\label{autocross_space}
	c(\chi)=\Bigg | \int_0^{Nd} e^{{\jmath}k\left(\frac{1}{2}p_1x^2+p_2x\right)}e^{{-\jmath}k\left(\frac{1}{2}p_1(x-\chi)^2+p_2(x-\chi)\right)}{\rm d}x\Bigg|,
\end{equation}
as the spatial autocorrelation function.\footnote{Strictly speaking, this is the modulus of the autocorrelation function, and this paper does not strictly distinguish between these two concepts. The autocorrelation and cross-correlation functions in the later text refer to the corresponding modulus size.}\footnote{Formula (\ref{autocross_space}) adopts a continuous form to express a more general situation, and in practical applications, it can be transformed into the corresponding discrete form. \cite{DBLP:journals/ojcs/PizzoSM22,DBLP:journals/tsp/PizzoTSM22}}

{\itshape Proposition 2 (Near Field Second Equivalent Principle):}
The autocorrelation function of the near-field is mathematically equivalent to the cross-correlation function of the far-field. Specifically, when $c'(\Delta k_x)=c'(\chi)$ is satisfied, there is: $\Delta k_x=p_1\chi$.

{\itshape proof :}
Affected by the modulus function, formula (\ref{autocross_space}) undergoes modification as follows:
\begin{equation}
	\begin{split}
		c(\chi)&=\Bigg|\int_0^{Nd}e^{{\jmath}k\left(p_1x\chi-\frac{1}{2}p_1\chi^2+p_2\chi\right)}{\rm d}x\Bigg|\\
		&=\Bigg|\int_0^{Nd}e^{{\jmath}kp_1x\chi}{\rm d}x\Bigg|.
	\end{split}
\end{equation}
In practical scenarios, the array size is limited, and conditions $0<x<Nd$ and $0<x-\chi<Nd$ must be satisfied. Consequently, $x$ must adhere to the inequality $\chi<x<Nd$. The autocorrelation function is truncated within this range to derive $c'(\chi)$:
\begin{equation}\label{c'chi}
	c'(\chi)=\Bigg|\int_\chi^{Nd}e^{{\jmath}kp_1x\chi}{\rm d}x\Bigg|.
\end{equation}
This function bears mathematical resemblance to the cross-correlation form in the far-field. Therefore, we leverage established insights from the far-field by drawing parallels between the cross-correlation function in the far-field and the autocorrelation function in the near-field. In the far-field, the cross-correlation function of the signal space in the $k_x$ and $k_x-\Delta k_x$ directions is expressed as:
\begin{equation}
	\begin{split}
		c(\Delta k_x)&=\Bigg|\int_0^{Nd}e^{{\jmath}\left(k_x-(k_x-\Delta k_x)\right)x}{\rm d}x\Bigg|\\
		&=\Bigg|\int_0^{Nd}e^{{\jmath}\Delta k_x x}{\rm d}x\Bigg|
	\end{split}
\end{equation}
To ensure mathematical consistency with equation (\ref{c'chi}), we also limit the integration range of $c(\Delta k_x)$ to $\chi<x<Nd$, yielding:
\begin{equation}\label{c'kx}
	c'(\Delta k_x)=\Bigg|\int_\chi^{Nd}e^{{\jmath}\Delta k_x x}{\rm d}x\Bigg|.
\end{equation}
The complete congruence between equations (\ref{c'chi}) and (\ref{c'kx}) substantiates Proposition 2.
$\hfill\blacksquare$

Proposition 2 asserts that the autocorrelation function value derived from the near-field can be inserted into the far-field array factor function or power pattern to acquire the equivalent $\Delta k_x$. Subsequently, $p_1$ can be determined through $p_1=\frac{\Delta k_x}{\chi}$. Therefore, Proposition 2 is aptly named the comparison principle. In cases where the cross-correlation method is employed to solve for $p_1$, a coupling between $p_1$ and $p_2$ emerges \cite{DBLP:journals/tcom/CuiD22, DBLP:journals/dsp/LiuMGLZT20 }. As a result, the JAC algorithm can decouple $p_1$ and $p_2$ by leveraging this principle in autocorrelation.

The above elucidates the complete process of solving for $p_1$ using the autocorrelation method. Hereafter, we will deploy the cross-correlation method to ascertain $p_2$ based on the previously determined $p_1$. As $p_2$ represents spatial frequency, and the cross-correlation method is more adept at frequency determination than the autocorrelation method, it remains the preferred choice for uncovering $p_2$.

{\itshape Proposition 3 (Near Field Third Equivalent Principle) :} The cross-correlation function between two signal spaces sharing the same $p_1$ but differing in $p_2$ within the near-field is analogous to the cross-correlation between two signal spaces oriented in distinct directions within the far-field. If $c(\Delta p_2)=c(\Delta k_x)$ holds true, then we can establish $\Delta p_2=\Delta k_x$.

{\itshape proof:}
If we denote $c(\Delta p_2)$ as the cross-correlation function between two signals with the same $p_1$ but different $p_2$, it can be expressed as:
\begin{equation}
	\begin{split}
		c(\Delta p_2)&=\Bigg|\int_0^{Nd}e^{jk\left(\left(\frac{1}{2}p_1x^2+p_2x\right)-\left(\frac{1}{2}p_1x^2+(p_2-\Delta p_2)x\right)\right)}{\rm d}x\Bigg|\\
		&=\Bigg|\int_0^{Nd}e^{jk\Delta p_2 x}{\rm d}x\Bigg|.
	\end{split}
\end{equation}
It is evident that the mathematical form of $c(\Delta p_2)$ is identical to that of $c(\Delta k_x)$. Therefore, when $c(\Delta p_2)=c(\Delta k_x)$, the equality $\Delta p_2=\Delta k_x$ holds.
$\hfill\blacksquare$

The third equivalent principle in the near-field states that, once $p_1$ is determined, the problem of finding $p_2$ in the near-field transforms into a far-field problem. We refer to this proposition as the degeneration principle. Consequently, beam training schemes such as the DFT codebook \cite{DBLP:conf/pimrc/Wang0LHCSA22} and hierarchical codebooks designed for far-field scenarios can be applied to determine $p_2$ in this context. This paper will not delve into the detailed discussion of these schemes. The aforementioned three equivalent principles constitute the entire process of the JAC algorithm.

\section{Main Advantages and Application Scenarios of the JAC Algorithm}

\subsection{The processing principle of signal space in JAC scheme}
In this section, we commence by employing matrix theory to analyze the configurations of far-field and near-field subspaces. Subsequently, we elucidate the significance and challenges associated with spatial orthogonal decomposition in near-field scenarios. Finally, we present how the JAC algorithm accomplishes orthogonal decomposition in the near-field space.

The spatial frequency spectrum of the received signal, as given by Equation (\ref{compressed channel}), can be represented through $N$ Fourier series. This linear representation is amenable to a matrix description. Let $\bm{\mathcal{H}}$ denote the full space matrix:
\begin{equation}
	{\bm{\mathcal{H}}} = [{\bm{a}}^*_{\theta_1},{\bm{a}}^*_{\theta_2},\cdots,{\bm{a}}^*_{\theta_N}] \in \mathbb{C}^{N\times N}.
\end{equation}

In far-field scenario, the projection value of the channel on each dimension is:
\begin{equation}
	{\bm {\mathcal F}}=[{\mathcal F}_{\theta_1},{\mathcal F}_{\theta_2},\cdots, {\mathcal F}_{\theta_N}]^\top.
\end{equation}
Therefore, the channel can be represented as:
\begin{equation}
	{\bm h}_{\rm far}={\bm {\mathcal F}}{\bm {\mathcal H}}.
\end{equation}
Assuming the set of numbers much greater than 0 in $\bm{\mathcal{F}}$ is denoted by $\bm{\Upsilon}$:
\begin{equation}
	\bm{\Upsilon} = \{\theta_1, \theta_2, \cdots, \theta_L\},
\end{equation}
where $L$ represents the number of multiple paths. Consequently, the channel can be approximated as follows:
\begin{equation}
	\bm{h}_{\text{far}} \approx \bm{\mathcal{F}}_{\bm{\Upsilon}} \bm{\mathcal{H}}_{:,\bm{\Upsilon}},
\end{equation}
Here, $\bm{\mathcal{H}}$ and $\bm{\mathcal{F}}$ can be regarded as a set of orthogonal bases and a linear combination of these orthogonal bases, respectively. Consequently, the far-field channel manifests as a low-dimensional hyperplane within a high-dimensional space.

In the near-field scenario, we initially consider the case where only a line of sight (LoS) path is present, with the assumption that the angle of arrival (AoA) for the LoS path at the reference antenna is $\theta_{\text{LoS}}$. The spatial Doppler shift part in the near-field is represented by $\bm{s}$ using the equation:
\begin{equation}\label{shaping vector}
	\bm{s}[n] = e^{\jmath k\frac{1}{2}p_1x^2_n}.
\end{equation}
Consequently, the channel can be expressed as:
\begin{equation}
	\bm{h}_{\text{near}} = \rho \bm{s} \cdot \bm{\mathcal{H}}_{:,\theta_{\text{LoS}}},
\end{equation}
where $\bm{s}$ is the vector determining the shape of the signal space, referred to as the shaping vector, and $\rho$ is the channel gain of the LoS path.  The signal subspace in the near-field exhibits a low-dimensional trend. Therefore, the fundamental difference between the near-field and far-field signal subspaces lies in the linearity of the far-field subspace versus the nonlinearity of the near-field subspace. Mathematically, the near-field signal subspace is jointly determined by the steering vector $\bm{\mathcal{H}}_{:,\theta_{\text{LoS}}}$ and the shaping vector $\bm{s}$. Unfortunately, the high degree of freedom associated with the shaping vector poses a challenge in designing an orthogonal codebook. This difficulty in near-field orthogonal decomposition arises not only due to the high freedom of the shaping vector but also because the shaping vector $\bm{s}$ and the steering vector $\bm{\mathcal{H}}_{:,\theta_{\text{LoS}}}$ are coupled together. Consequently, we can only obtain the projection of the signal space on the low-dimensional fashion determined jointly by the shaping vector and steering vector.

Addressing the challenge of near-field signal space orthogonal decomposition, many studies have resorted to spatial approximate orthogonal decomposition schemes for beam training design. However, this approach renders common beam training performance metrics, such as average achievable rate and mismatch ratio, ineffective. The critical factor is that only when space's orthogonal decomposition is achieved can we ensure that, irrespective of the signal's location, there is always a codeword in the codebook capable of elevating the user's received power beyond the threshold. In such cases, the primary variable influencing beam training accuracy and communication signal quality becomes noise, and average performance metrics adequately measure the algorithm's quality.

Conversely, when approximate orthogonal decomposition of space is the best that can be achieved, there will inevitably be regions left uncovered by the codebook. In these uncovered areas, the primary determinant of communication performance is not the magnitude of channel noise but rather the design of the codebooks.\footnote{To avoid leaving any portion of space unaccounted for, the codebook's number of codewords needs to be increased, achieving coverage of all regions through repetitive searches. In essence, if orthogonality between codewords cannot be guaranteed, any region covered by codewords in the codebook without duplication will experience omissions. Conversely, ensuring no omissions will result in duplicated regions. Therefore, the relationship between spatial coverage and the time complexity of the codebook becomes a trade-off, a consideration not applicable in far-field codebook design and beam training scenarios.} Consequently, communication errors for users in these regions are non-random and cannot be effectively gauged using performance indicators with average significance.

The JAC algorithm achieves orthogonal decomposition in the near-field space. Let $\bm{\widetilde h}$ denote the feature tangent plane of the low dimensionality of the signal subspace. This plane, situated at the reference antenna, acts as the tangent plane of the signal subspace. Because this hyperplane encapsulates information about the channel's steering vector, we refer to it as the feature hyperplane of the channel. The feature tangent plane of the near-field signal subspace exclusively retains the dimensional features of the signal space, omitting the shape features. The choice of tangent points on the feature tangent plane does not impact our analysis of the near-field signal subspace, and we only need to select a fixed tangent point. This paper designates the tangent plane of the signal space at the reference antenna as the characteristic tangent plane of the near-field signal subspace. We define $\bm{\widetilde h}$ as:
\begin{equation}
	{\bm{\widetilde h}} = \rho{\bm s}^* \cdot {\bm h}_{\rm near} = \rho{\bm{\mathcal H}}_{:,\theta_{\rm LoS}},
\end{equation}
which represents a hyperplane rather than a trend.\footnote{In a scenario with only a Line of Sight (LoS) path, the signal subspace is one-dimensional and does not strictly qualify as a hyperplane. However, this paper emphasizes linearity or nonlinearity rather than the dimensionality of space. Therefore, linear spaces are collectively referred to as hyperplanes, while nonlinear spaces are collectively referred to as trends.}

The feature tangent plane of the near-field signal subspace aligns with the form of the far-field signal subspace. Leveraging the far-field analysis method, we can analyze the feature tangent plane of the near-field signal space using the Fourier series to represent the channel gain $\rho$. This leads to the following representation:
\begin{equation}\label{hyperplane}
	{\bm{\widetilde h}} = {\bm{\mathcal F}}_{\theta_{\rm LoS}} \cdot {\bm{\mathcal H}}_{\theta_{\rm LoS}}.
\end{equation}
Here, the coefficients of the Fourier series can be expressed as:
\begin{equation}\label{nearfou}
	\begin{split}
		{\bm{\mathcal F}}_{\theta_{\rm LoS}} &= \frac{1}{N} \left| \sum_{n=1}^{N} {\bm a}_{\theta_{\rm LoS}}[n] {\bm s}^*[n] {\bm r}[n] \right| \\
		&= \frac{1}{N} \left| \sum_{n=1}^N {\bm b}_{\theta_{\rm LoS}}[n] {\bm r}[n] \right|,
	\end{split}
\end{equation}
where ${\bm b}_{\theta_{\rm LoS}}$ is the near-field beam focusing vector \cite{DBLP:journals/tcom/LuD23, DBLP:journals/cm/ZhangSGDE23, 7912361}, and ${\bm a}_{\theta_{\rm LoS}}$ is the steering vector of the near-field feature tangent plane. It's noteworthy that the near-field beam focusing vector is jointly determined by the shaping vector and the steering vector. By utilizing equation (\ref{nearfou}), the near-field channel can be redefined as:
\begin{equation}\label{near_channel_fou}
	{\bm h}_{\rm near} = {\bm s} \cdot \left( {\bm{\mathcal F}}_{\theta_{\rm LoS}}{\bm H}_{:,\theta_{\rm LoS}} \right).
\end{equation}
Examining equation (\ref{near_channel_fou}), we observe that by first determining the shaping vector, we can transform the near-field problem into the same form as the far-field problem, facilitating orthogonal decomposition. The pivotal question then becomes whether we can directly ascertain the shaping vector of the channel without knowledge of the steering vector.

In accordance with equation (\ref{shaping vector}), it is evident that the shaping vector is entirely determined by $p_1$. Simultaneously, the autocorrelation function of the signal subspace is solely a function of $p_1$:
\begin{equation}
	\begin{split}
		c(\nu) &= \Bigg| \sum_{n=1}^{N-\nu} h[n] h^*[n+\nu]\Bigg|\\
		&= \rho^2 \Bigg| \sum_{n=1}^{N-\nu} s[n] s^*[n+\nu]\Bigg|\\
		&= \rho^2 \Bigg| \sum_{n=1}^{N-\nu} s[n]s[\nu] s^*[n+\nu]\Bigg|\\
		&= \rho^2 \Bigg| \sum_{n=1}^{N-\nu}e^{jp_1\nu nd}\Bigg|,
	\end{split}
\end{equation}
where $\nu$ is an integer. Consequently, the shaping vector can be determined through the autocorrelation function, effectively solving the shape of the signal space via autocorrelation. Subsequently, using equation (\ref{hyperplane}), we can ascertain the feature hyperplane corresponding to the signal subspace. The near-field signal subspace post-projection aligns entirely with the far-field signal subspace, allowing the utilization of far-field methods to determine the near-field signal subspace post-projection. The intricacies of the proposed algorithm are summarized in Algorithm 1.

\begin{algorithm}
	
	\caption{JAC beam training scheme }
	\begin{algorithmic}[1]\label{JAC_alg}
		\STATE {Pre select the threshold $\eta$ and calculate the value of $\Delta k_x$ when $c(\Delta k_x)=\eta$. Prepare DFT codebook $\bm \Theta$ based on the number of antennas of the array.  }
		
		\STATE{Receive the signal $\bm r$ sent by the user.}
		\STATE{Use equation (\ref{autocross_space}) to solve the autocorrelation function $c(\chi)$. }
		\STATE{Find $\chi_0$ at $c(\chi_0)=\eta$ based on the pre-set threshold $\eta$.}
		\STATE{According to the second equivalent principle in the near-field, obtain $p_1=\frac{\Delta k_x}{\chi_0}$.}
		\STATE{According to equation (\ref{shaping vector}), obtain the shaping vector $\bm s$.}
		\STATE{Multiply the shaping vector $\bm s$ with the DFT codebook $\bm \Theta$ to obtain a new near-field codebook $ \widetilde{\bm \Theta}$.}
		\STATE{Using  $ \widetilde{\bm \Theta}$ for beam training, the beam corresponding to the maximum received power by the user is the optimal beam.}

	\end{algorithmic}
\end{algorithm}
\vspace{-0in}%

\subsection{Analysis of Time Complexity of JAC Algorithm}

\begin{table*}[!htbp]\label{time_complexity}
	\centering
	\caption{\textsc{Time Complexity of Near Field Algorithms}}
	\begin{tabularx}{\linewidth}{|p{2.5cm}|p{12.00cm}|p{2.34cm}|}
		\hline
		\textbf{ Algorithm name } &  \textbf{Algorithm principle } & \textbf{Time complexity  }\\
		\hline
		{Polar codebook } & {Create a 2D codebook in both distance and direction.  Design based on the orthogonality principle between adjacent codewords in terms of distance. Can be applied in both the far and near-fields, and automatically degenerates into a DFT codebook in the far-field.} & {$\mathcal{O}(SN)$ }\\
		\hline
		{DFT codebook } &  {Completely disregard CoA. Only design codebooks in the direction. Is suitable for far-field applications, with significant power loss in the near-field.} & {$\mathcal{O}(N)$ }\\
		\hline
		{Proposed codebook } &  {Use autocorrelation method on CoA estimation and DFT codebook on AoA estimation. Can be applied in both the far and near-fields, and automatically degenerates into a DFT codebook in the far-field. } & {$\mathcal{O}(N+1)$ }\\

		\hline
	\end{tabularx}%
	\label{t2}%
	
\end{table*}%
It is crucial to clarify that the time complexity mentioned here refers to the number of basic beams required to cover the entire near-field space. In practical applications, the JAC algorithm can be combined with other algorithms designed to reduce time complexity in the far-field further. For example, methods leveraging beam splitting phenomena in broadband scenarios or utilizing hierarchical codebooks can be integrated with the JAC algorithm to achieve enhanced time efficiency \cite{DBLP:journals/twc/CuiDWZG23}. When comparing time complexity, to maintain fairness, we focus on the time complexity when using the basic beam, as different forms of beam training methods can be combined with various approaches to reduce time complexity. Table II presents the complexity of the polar codebook, DFT codebook, and the proposed JAC codebook.

For the polar codebook, a two-dimensional search in distance and angle is required in the near-field space. Regarding angle, the search frequency of the polar codebook aligns with that of the DFT codebook, leading the polar codebook to automatically degenerate into the form of the DFT codebook in the far-field. Concerning distance, the sampling number of the polar codebook is denoted as $S$, where $S$ represents the sampling number on the distance and is positively correlated with $N$. The use of $\mathcal{O} (SN)$ instead of $\mathcal{O}(N^2)$ in Table II is due to the fact that $S$ is generally much smaller than $N$. However, it's essential to note that even though $S$ is much smaller than $N$, the size of the polar codebook is still considerably larger than that of the DFT codebook, especially evident in XL-MIMO scenarios where the volume of the codebook is determined by their multiplication.

Including the DFT codebook in this table is for the sole purpose of comparing time complexity. The DFT codebook is limited to far-field use and cannot be applied in the near-field, where it sacrifices beamforming gain.

Our proposed JAC codebook scheme does not significantly increase time complexity compared to DFT codebooks, yet it accomplishes CoA estimation and can be effectively employed in near-field scenarios. The JAC algorithm consists of two steps: firstly, using the autocorrelation method to solve $p_1$, and secondly, using the cross-correlation method to solve $p_2$. As the JAC algorithm does not consume time resources when determining $p_1$, the time complexity of the first stage is $\mathcal O (1)$, independent of array size and communication distance. The beam training process in the second stage is entirely consistent with the far-field beam training process, resulting in a time complexity of $\mathcal O (N)$. Therefore, the overall time complexity of the JAC algorithm is $\mathcal O (N+1)$.

\subsection{Application scenarios of JAC algorithm}
The JAC algorithm is well-suited for digital beamforming scenarios in XL-MIMO systems. In practical applications, the received signal is discretized in space, meaning that signals are only received at the antenna positions. Consequently, when performing autocorrelation, only discrete autocorrelation function values can be obtained. The discreteness of the autocorrelation function introduces challenges, as it may not always be possible to find an exact match for the value of $\chi_0(\eta)$ that equals $\eta$. This imprecision in $\chi_0(\eta)$ can lead to errors in the estimation of $p_1$.

Autocorrelation is more susceptible to noise compared to cross-correlation, primarily due to the multiplicative effect of noise from different antennas during the autocorrelation function calculation. Moreover, when solving the autocorrelation function, the array does not employ any combining measures to enhance the Signal-to-Noise Ratio (SNR) of the received signal. This lack of combining measures can result in a more substantial estimation error of $p_1$, particularly in low SNR scenarios. Consequently, the JAC scheme is better suited for beam training or user localization problems in scenarios characterized by high SNR.

To enhance the accuracy of $p_1$ estimation, the JAC algorithm is designed with a preference for arrays adopting digital beamforming. This design choice enables the independent extraction of the received signal from each antenna, contributing to more accurate parameter estimation.

\section{Simulation Results}
In this section, we present numerical results to validate the efficacy of the proposed algorithms. Our focus is on the uplink channel beam training scenario, where the base station (BS) is a Uniform Linear Array (ULA) with $N$ antennas and a size of $D$, while the user is equipped with a single antenna. For our simulations, we set $N=800$ and $D=2$, with an antenna spacing distance of $\frac{\lambda}{2}$ and a signal frequency of 60 GHz. In this configuration, the Rayleigh distance is $\frac{2D^2}{\lambda}=1600$ m, ensuring that the user consistently resides in the near-field region of the BS throughout our simulations.

We assume an environment with no other scatterers, implying a Line-of-Sight (LoS) path as the sole communication channel between the user and the BS. The beamforming vector on the BS side is denoted as ${\bm \omega} \in \mathbb{C}^{N \times 1}$. To facilitate a fair comparison of different beamforming schemes, we disregard signal attenuation during the propagation process. The Signal-to-Noise Ratio (SNR) is defined as:
\begin{equation}
	SNR = \frac{P_t N}{\sigma^2},
\end{equation}
where $P_t$ represents the transmission power. The achievable rate is given by:
\begin{equation}
	R = \log_2 \left(1 + \frac{P_t N \vert {\bm \omega}^{\top}{\bm h}\vert^2}{\sigma^2}\right),
\end{equation}
and the simulation results are averaged over 100 randomly distributed users.

To illustrate the spatial coverage of various codebook schemes in the absence of noise, we define the maximum radiation power at position $\bm{x}$ as
\begin{equation}
	P_r(\bm{p}) = P_t N \vert {\bm{\omega}}_0^\top {\bm{h}} \vert^2,
\end{equation}
where ${\bm{\omega}}_0$ is the optimal codeword at position $\bm{p}$. The ratio of $P_r$ to $P_t$ is defined as the coverage of the codebook at position $\bm{p}$:
\begin{equation}
	R_{\text{cover}} = \frac{P_r}{P_t} = N \vert {\bm{\omega}}_0^\top {\bm{h}} \vert^2.
\end{equation}
$R_{\text{cover}}$ represents the capability of different codebook schemes to project signal power to users at various locations in a noiseless channel. For a comparative analysis, the proposed codebook schemes are compared with the following schemes:

\begin{itemize}
	\item Polar codebook: The codebook is partitioned into grids using both angle and distance variables. Angular partitioning is uniform, similar to the DFT codebook, while distance partitioning is non-uniform to adhere to the orthogonality principle between adjacent codewords.

	\item DFT codebook: The codebook evenly divides space only based on angles, neglecting distance variables. While this codebook serves as a far-field reference, it effectively sets a performance baseline for near-field codebooks. By contrasting it with DFT codebooks, the extent of performance enhancement achievable through the introduction of CoA in the near-field becomes evident.

\end{itemize}
\subsection{Average Achievable  Rate}
\begin{figure} 
	\centering
	\includegraphics[width=\linewidth]{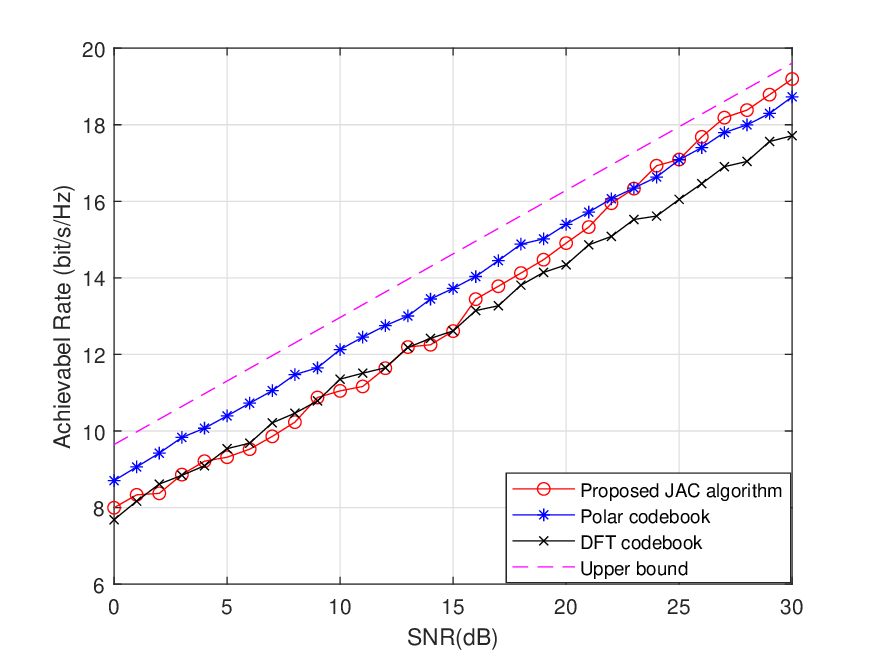}
	\caption{ Achievable rate against SNR for several codebook schemes in ULA channel, with N = 800, f = 60 GHz and communication distance is (5,100) m.}   \label{f1} 
\end{figure}
In Fig. \ref{f1}, the average achievable rates for the three schemes are depicted for SNR values ranging from 0 to 30 dB and distances from the user to the reference antenna of the BS ranging from 5 to 100 m. The pink dashed line represents the achievable rate corresponding to the upper bound of beamforming, i.e., $\vert {\bm \omega}^\top {\bm h} \vert^2=1$. Notably, when the SNR is below 15 dB, the JAC scheme's performance aligns closely with that of the DFT codebook, suggesting challenges in accurately estimating $p_1$ using the autocorrelation method, impacting the JAC scheme's operation. For SNRs between 15 and 25 dB, the JAC algorithm significantly outperforms the DFT codebook in average achievable rate, with comparable beam training time. Despite not matching the polar codebook's average achievable rate, the JAC algorithm exhibits remarkable overall performance. As the SNR exceeds 25 dB, the JAC scheme surpasses the polar codebook in achievable rate. Considering its low time complexity, the JAC scheme demonstrates a clear performance advantage.

\begin{figure}
	\centering
	\includegraphics[width=\linewidth]{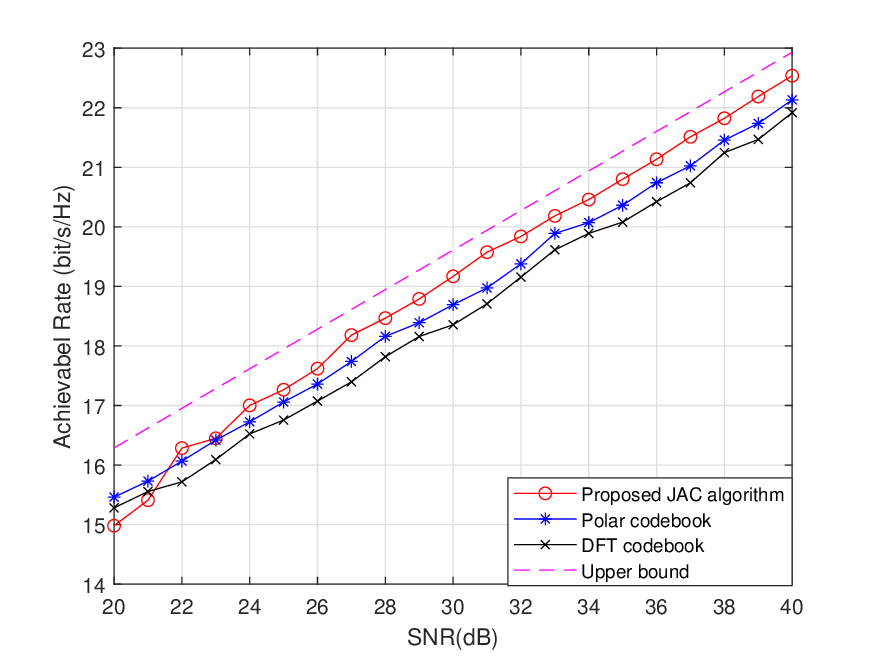}
	\caption{ Achievable rate against SNR for several codebook schemes in ULA channel, with N = 800, f = 60 GHz and communication distance is (5,200) m.}     \label{f3}
\end{figure}
In Fig. \ref{f3}, the average achievable rates for the three schemes are illustrated for distances from the user to the reference antenna of the BS ranging from 5 to 200 m and SNR values from 20 to 40 dB. Notably, when the SNR is 20 dB, the JAC scheme exhibits a lower average achievable rate than the DFT codebook. This contrasts with Fig. \ref{f1}, where the JAC scheme outperformed the DFT codebook at an SNR of 20 dB. The discrepancy arises due to the inverse relationship between distance from the user to the array and the accuracy of $p_1$ estimation under the same SNR, leading to diminished JAC algorithm performance. However, for SNRs exceeding 25 dB, the JAC scheme's performance once again surpasses that of both the DFT codebook and polar codebook. As the SNR increases, the JAC scheme gradually approaches the upper bound, indicating that further increases in SNR beyond a certain threshold have limited impact on the JAC scheme's performance.

\subsection{Space Coverage Rate}
\begin{figure}
	\centering
	\includegraphics[width=\linewidth]{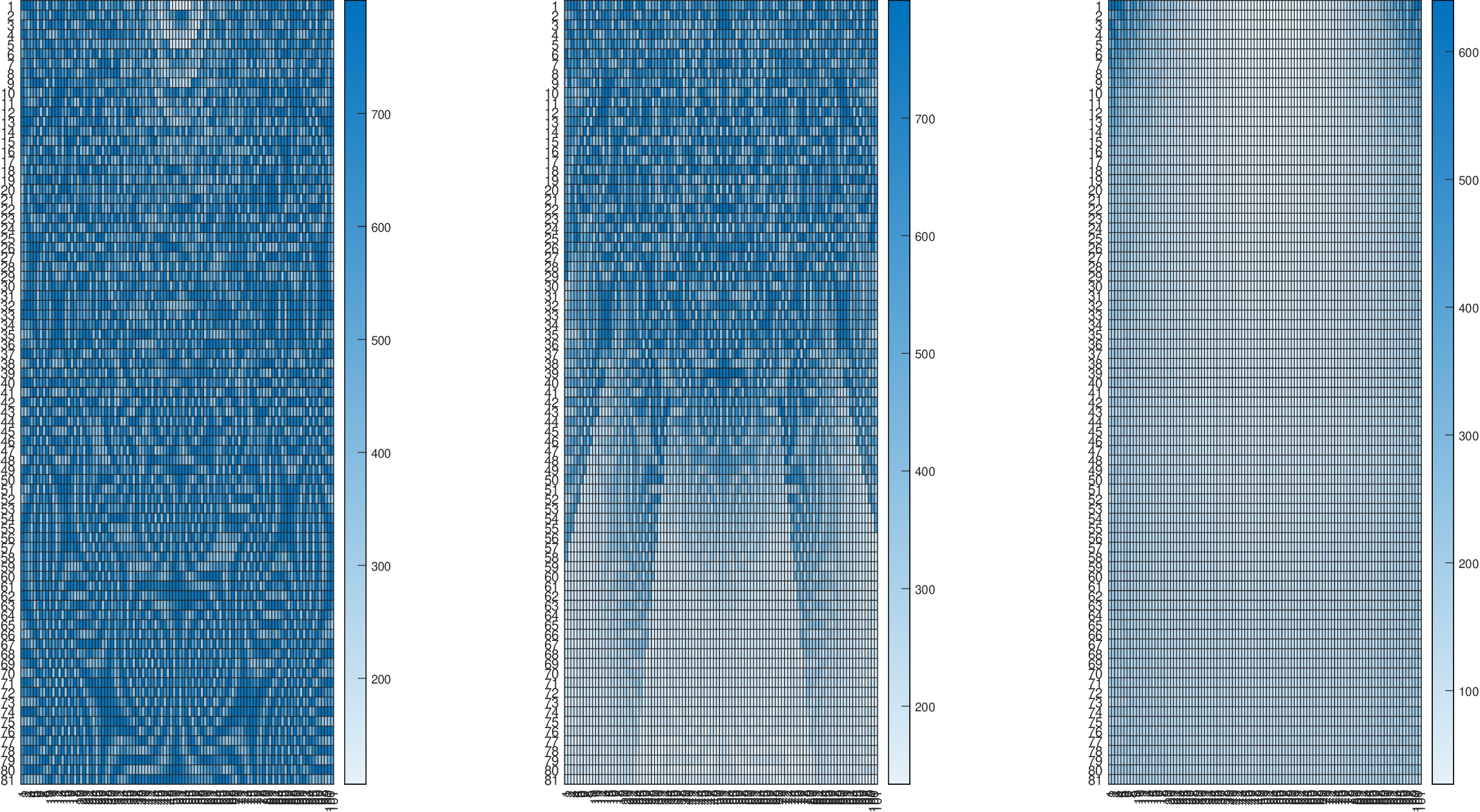}
	\caption{ Heatmap of several codebook schemes with the space of $x\in(20,100)$ and $z\in (-50,50)$.}     \label{heat1}
\end{figure}

Following the experiment on average achievable rates, it was observed that none of the three schemes could fully match the upper bound. Consequently, we conducted experiments to compare the coverage of the three schemes at various locations in space in a noiseless environment.

Fig. \ref{heat1} illustrates the coverage across the range $x \in (20,100)$ and $z \in (-50,50)$. Notably, at close distances, the spatial coverage of the JAC scheme and the polar codebook appears similar. However, as the distance increases, the coverage of the polar codebook significantly diminishes. Simultaneously, both the JAC scheme and the polar codebook exhibit much higher coverage than the DFT codebook. The reduction in coverage for polar codebooks can be attributed to their design, which primarily considers the orthogonality between two codewords. However, the spatial range corresponding to one codeword is not necessarily strongly correlated. In the polar codebook, codewords are formed based on a specific point in space, resulting in insufficient coverage for positions unrelated to this point. Furthermore, in the polar codebook, as the distance decreases, the grid division of codewords becomes denser, reducing the likelihood of insufficient coverage. Conversely, for relatively distant positions in the near-field, a significant portion experiences insufficient coverage.

\begin{figure}
	\centering
	\includegraphics[width=\linewidth]{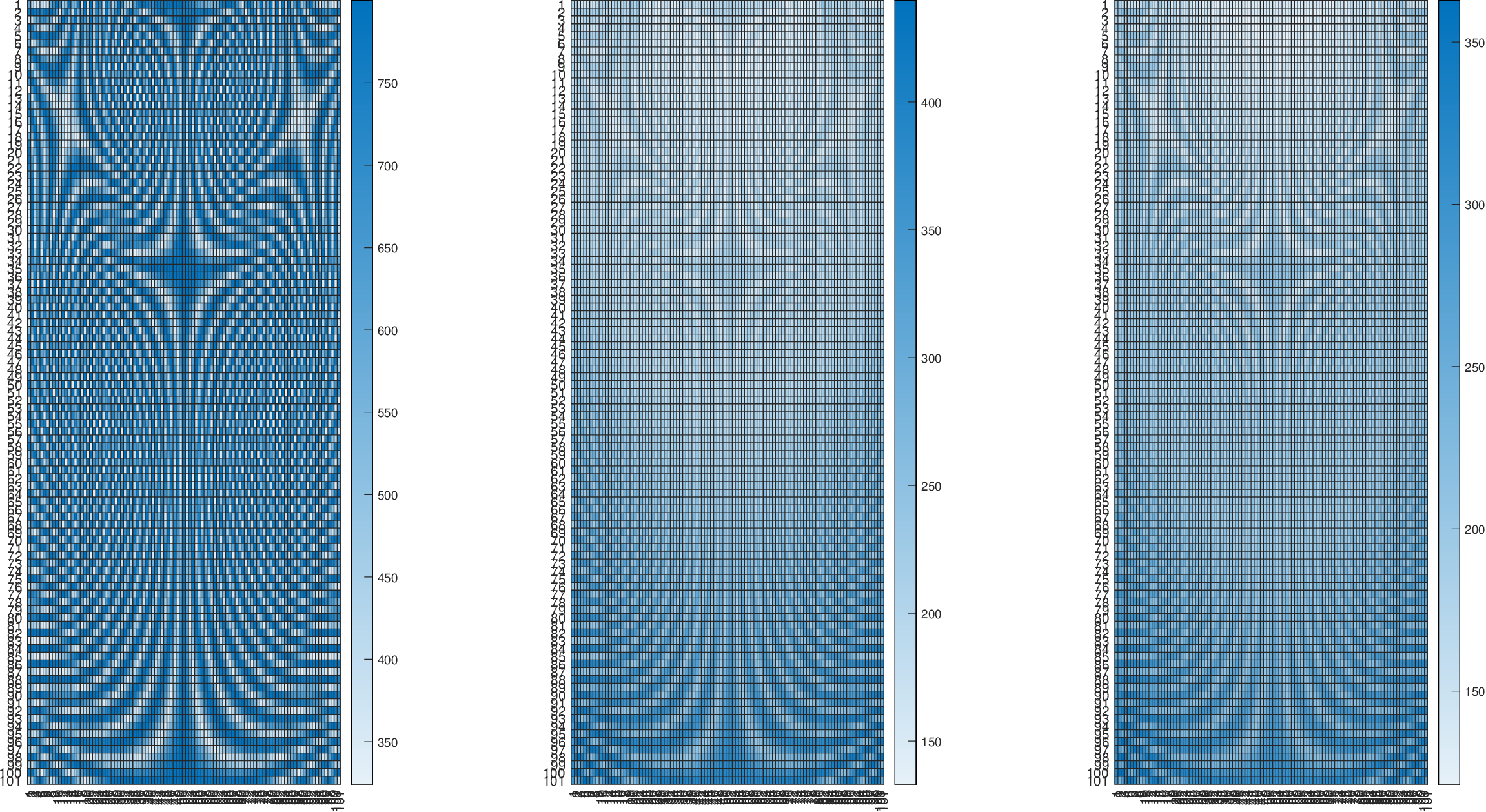}
	\caption{ Heatmap of several codebook schemes with the space of $x\in(100,200)$ and $z\in (-50,50)$.}     \label{heat2}
\end{figure}

Fig. \ref{heat2} displays the coverage across the range $x \in (100,200)$ and $z \in (-50,50)$. Notably, in this range, the coverage of the JAC scheme outperforms the other two schemes significantly. Interestingly, all three schemes exhibit the same pattern shape in this range. This similarity arises because both the JAC solution and the polar codebook are implemented by adding CoA information on top of the DFT codebook. Additionally, it's worth noting that in the pattern of the JAC scheme, the coverage corresponding to the lightest color position is approximately 400, indicating $\vert {\bm \omega}_0 \vert^2 = 0.5$. This suggests that the JAC scheme can ensure that almost all positions in the near-field experience a maximum power loss of 3 dB under high SNR conditions.

\begin{figure}
	\centering
	\includegraphics[width=\linewidth]{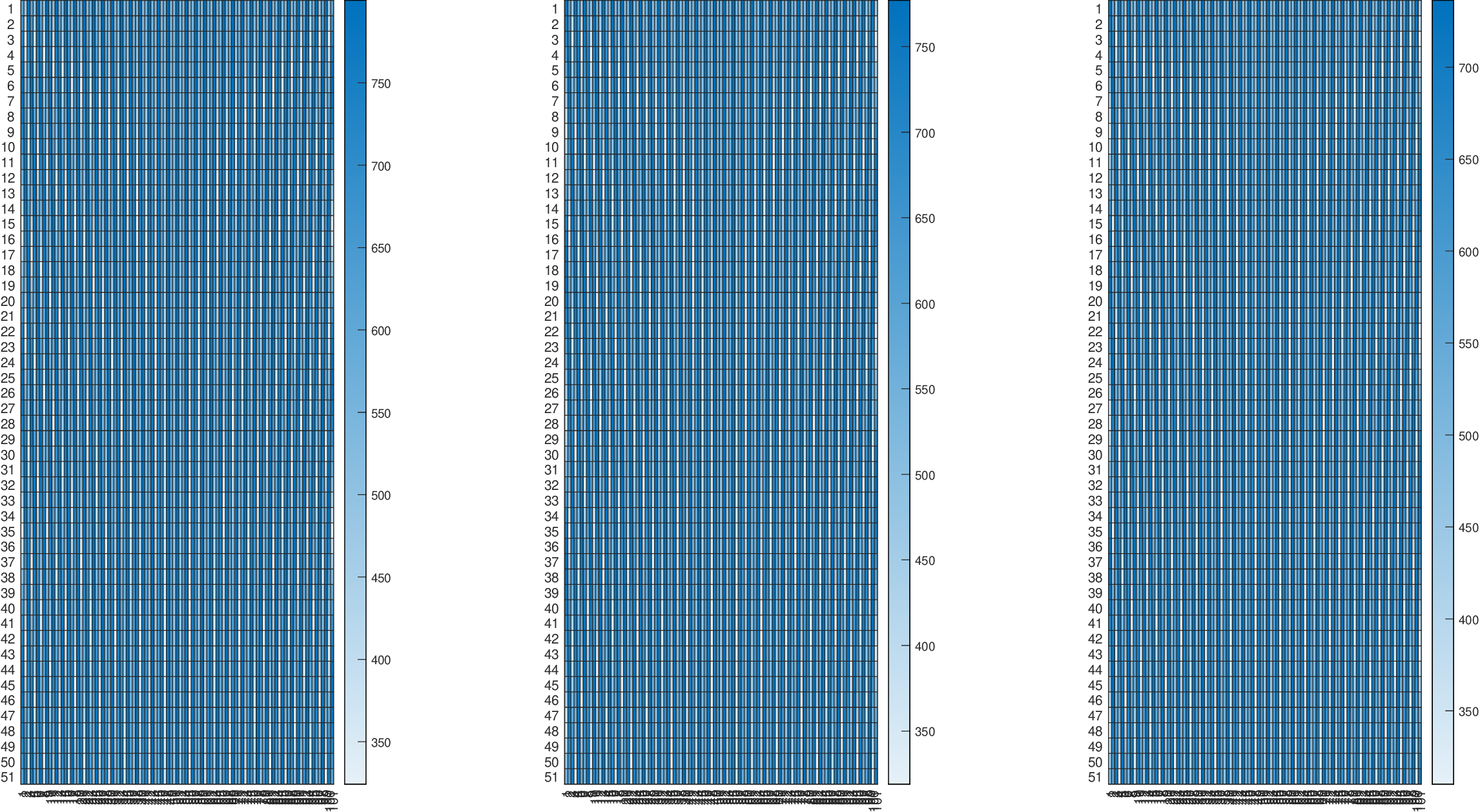}
	\caption{ Heatmap of several codebook schemes with the space of $x\in(600,650)$ and $z\in (-50,50)$.}     \label{heat3}
\end{figure}

Fig. \ref{heat3} depicts the coverage across the range $x \in (600,650)$ and $z \in (-50,50)$. In this range, we observe that the pattern shapes of the three schemes are approximately similar. The JAC scheme performs slightly better than the polar codebook, while the polar codebook is slightly better than the DFT codebook. This observation suggests that both the JAC scheme and the polar codebook naturally transition to the DFT codebook in the far-field.

\section{Conclusion}
This paper mainly proposed a scheme to reduce the time complexity of near-field beam training in high SNR scenarios.  Specifically, the problem of near-field beam training in this paper were seen as the estimation of the coefficients of the quadratic and linear terms of the near-field channel phase in space. Based on detailed formula derivation, we found that the coefficients of the quadratic term can be determined using autocorrelation without knowing the coefficients of the linear term. When the quadratic coefficient were determined, the problem of solving the linear coefficient were mathematically consistent with the beam training problem in the far-field. Therefore, far-field beam training schemes could be used to complete the beam search process at this stage. The proposed JAC scheme saved a lot of beam training time compared to other near-field codebooks, and the time complexity of near-field beam training were reduced from $\mathcal{O}(N^2)$ to $\mathcal{O}(N+1)$. At a high SNR regime, the performance of the JAC algorithm in terms of spatial coverage and average reachability rate were more prominent. However, in low SNR regime, the performance of the JAC algorithm decreases because autocorrelation is more susceptible to noise interference compared to cross correlation. We will further improve the noise resistance performance of the JAC algorithm in future work.

\ifCLASSOPTIONcaptionsoff
\newpage
\fi



%

\appendices

\section{proof of proposition 1}\label{p1proof}
Assuming ${\bm r}_{\rm dsp}(t)$ is the received signal in the DSP and ${\bm r}_{\rm asp}(x)$ is the received signal in the ASP. In the absence of time-domain Doppler and spatial Doppler, the signal forms of both are:
\begin{equation}
	{\bm r}_{\rm dsp}(t)=e^{{\jmath}\omega t}
\end{equation}
and 
\begin{equation}
	{\bm r}_{\rm asp}(x)=e^{{\jmath}k_x x},
\end{equation}
where $k_x=ksin\theta$ is the spacial frenquency along $x$ axis. When the user is in motion or near-field, the time-domain frequency and spatial frequency of the received signal in the DSP and ASP are functions of time and space, respectively, which can be written as:
\begin{equation}
	{\bm r}_{\rm dsp}(t)=e^{{\jmath}\omega(t)t}
\end{equation} 
and
\begin{equation}
	{\bm r}_{\rm asp}(x)=e^{{\jmath}k_x(x) x}.
\end{equation}
In DSP, $\vert \Delta \omega(t)\vert_{\rm max}$ is the maximum Doppler frequency offset.  Imitating this concept, we define the maximum Doppler direction shift in ASP as $\vert \Delta k_x(x)\vert_{\rm max}$. According to formula (\ref{sim_near}), there is $k_x(x)=p_1 x+p_2$. Thus $\vert \Delta k_x(x)\vert_{\rm max}=p_1 Nd$ holds. Since $N$ and $d$ are known variables, as long as $\vert \Delta k_x(x)\vert_{\rm max}$ is estimated, $p_1$ can be known. In this way, we turn the problem of parameter estimation in the near-field into the problem of estimating the maximum Doppler in signal processing.  In DSP, coherence time is the reciprocal of the maximum Doppler frequency shift:
\begin{equation}
	\tau_0=\frac{1}{\vert \Delta \omega(t)\vert_{\rm max}},
\end{equation}
where $\tau_0$ is coherence time. By analogy with the concept of coherent time, we propose the concept of coherent space in ASP:
\begin{equation}\label{chi0}
	\chi_0=\frac{1}{\vert \Delta k_x(x)\vert_{\rm max}},
\end{equation}
where $\chi_0$ is the coherent space. In some papers, XL-MIMO is divided into different tiles, and the channel can be viewed as far-field for each tile. The principle behind this is that the spatial size of a tile is smaller than the coherent space of the channel \cite{DBLP:conf/vtc/LiWHCS22}. Furthermore, when solving the coherent space, we can artificially specify a threshold $\eta$ of coherence as a reference. If the coherence of two signals is greater than $\eta$, we consider them coherent; if it is less than $\eta$, we consider them independent of each other. Because the definition of Rayleigh distance varies in different situations \cite{DBLP:conf/acssc/BjornsonDS21}, we call $\chi_0(\eta)$ the coherent space under the $\eta$ criterion.  According to formula (\ref{chi0}), there is:
\begin{equation}
	\vert \Delta k_x(x)\vert_{\rm max}=\alpha(\eta)\frac{1}{\chi_0(\eta)},
\end{equation}
where $\alpha(\eta)$ is a constant independent of $x$ and $k_x$. Therefore, we can obtain:
\begin{equation}\label{p1&cospace}
	p_1Nd=\alpha(\eta)\frac{1}{\chi_0(\eta)}.
\end{equation}
In equation (\ref{p1&cospace}), $N$ and $d$ are known constants, and $\alpha(\eta)$ is the coefficient generated based on the selected threshold $\eta$, so the final variable $p_1$ is only a function of $\chi_0(\eta)$. Therefore, the process of solving $p_1$ can be transformed into the process of solving coherent space size, i.e. $\chi_0(\eta)$. Therefore, $p_1$ and $p_2$ can be decoupled. The proposition is proved.

%






\bibliographystyle{IEEEtran} 
\bibliography{myref}  

\begin{thebibliography}{10}
\providecommand{\url}[1]{#1}
\csname url@samestyle\endcsname
\providecommand{\newblock}{\relax}
\providecommand{\bibinfo}[2]{#2}
\providecommand{\BIBentrySTDinterwordspacing}{\spaceskip=0pt\relax}
\providecommand{\BIBentryALTinterwordstretchfactor}{4}
\providecommand{\BIBentryALTinterwordspacing}{\spaceskip=\fontdimen2\font plus
\BIBentryALTinterwordstretchfactor\fontdimen3\font minus \fontdimen4\font\relax}
\providecommand{\BIBforeignlanguage}[2]{{%
\expandafter\ifx\csname l@#1\endcsname\relax
\typeout{** WARNING: IEEEtran.bst: No hyphenation pattern has been}%
\typeout{** loaded for the language `#1'. Using the pattern for}%
\typeout{** the default language instead.}%
\else
\language=\csname l@#1\endcsname
\fi
#2}}
\providecommand{\BIBdecl}{\relax}
\BIBdecl

\bibitem{DBLP:journals/bell/Marzetta15}
T.~L. Marzetta, ``Massive {MIMO:} an introduction,'' \emph{Bell Labs Tech. J.}, vol.~20, pp. 11--22, Mar. 2015.

\bibitem{DBLP:journals/twc/ZhengZ23}
B.~Zheng and R.~Zhang, ``Simultaneous transmit diversity and passive beamforming with large-scale intelligent reflecting surface,'' \emph{{IEEE} Trans. Wirel. Commun.}, vol.~22, no.~2, pp. 920--933, Feb. 2023.

\bibitem{DBLP:journals/twc/WuZ19}
Q.~Wu and R.~Zhang, ``Intelligent reflecting surface enhanced wireless network via joint active and passive beamforming,'' \emph{{IEEE} Trans. Wirel. Commun.}, vol.~18, no.~11, pp. 5394--5409, Nov. 2019.

\bibitem{DBLP:conf/iccchina/PengZJLL23}
X.~Peng, L.~Zhao, Y.~Jiang, J.~Liu, and W.~Li, ``Channel estimation for extremely large-scale massive {MIMO} systems in hybrid-field channel,'' in \emph{{IEEE/CIC} International Conference on Communications in China, {ICCC} 2023, Dalian, China, August 10-12, 2023}.\hskip 1em plus 0.5em minus 0.4em\relax {IEEE}, Aug. 2023, pp. 1--6.

\bibitem{DBLP:conf/acssc/TorresSB20}
A.~de~Jesus~Torres, L.~Sanguinetti, and E.~Bj{\"{o}}rnson, ``Near- and far-field communications with large intelligent surfaces,'' in \emph{54th Asilomar Conference on Signals, Systems, and Computers, {ACSCC} 2020, Pacific Grove, CA, USA, November 1-4, 2020}, M.~B. Matthews, Ed.\hskip 1em plus 0.5em minus 0.4em\relax {IEEE}, Nov. 2020, pp. 564--568.

\bibitem{DBLP:conf/wcsp/WangZZTY21}
T.~Wang, K.~Zhang, Y.~Zhang, H.~Tong, and C.~Yin, ``Near-field beam management in lis-assisted mmwave systems,'' in \emph{13th International Conference on Wireless Communications and Signal Processing, {WCSP} 2021, Changsha, China, October 20-22, 2021}.\hskip 1em plus 0.5em minus 0.4em\relax {IEEE}, Oct. 2021, pp. 1--6.

\bibitem{DBLP:conf/pimrc/HuIW22}
S.~Hu, M.~C. Ilter, and H.~Wang, ``Near-field beamforming for large intelligent surfaces,'' in \emph{2022 {IEEE} 33rd Annual International Symposium on Personal, Indoor and Mobile Radio Communications (PIMRC), Kyoto, Japan, September 12-15, 2022}.\hskip 1em plus 0.5em minus 0.4em\relax {IEEE}, Jun. 2022, pp. 1367--1373.

\bibitem{10146329}
S.~Hu, H.~Wang, and M.~C. Ilter, ``Design of near-field beamforming for large intelligent surfaces,'' \emph{IEEE Transactions on Wireless Communications}, Jun. 2023.

\bibitem{DBLP:journals/tsp/Friedlander19a}
B.~Friedlander, ``Localization of signals in the near-field of an antenna array,'' \emph{{IEEE} Trans. Signal Process.}, vol.~67, no.~15, pp. 3885--3893, Aug. 2019.

\bibitem{DBLP:journals/ojcs/BjornsonS20}
E.~Bj{\"{o}}rnson and L.~Sanguinetti, ``Power scaling laws and near-field behaviors of massive {MIMO} and intelligent reflecting surfaces,'' \emph{{IEEE} Open J. Commun. Soc.}, vol.~1, pp. 1306--1324, Sep. 2020.

\bibitem{DBLP:journals/tcom/CuiD22}
M.~Cui and L.~Dai, ``Channel estimation for extremely large-scale {MIMO:} far-field or near-field?'' \emph{{IEEE} Trans. Commun.}, vol.~70, no.~4, pp. 2663--2677, Apr. 2022.

\bibitem{7942128}
K.~T. Selvan and R.~Janaswamy, ``Fraunhofer and fresnel distances: Unified derivation for aperture antennas,'' \emph{IEEE Antennas and Propagation Magazine}, Aug. 2017.

\bibitem{DBLP:conf/acssc/BjornsonDS21}
E.~Bj{\"{o}}rnson, {\"{O}}.~T. Demir, and L.~Sanguinetti, ``A primer on near-field beamforming for arrays and reconfigurable intelligent surfaces,'' in \emph{55th Asilomar Conference on Signals, Systems, and Computers, {ACSSC} 2021, Pacific Grove, CA, USA, October 31 - November 3, 2021}.\hskip 1em plus 0.5em minus 0.4em\relax {IEEE}, Oct. 2021, pp. 105--112.

\bibitem{DBLP:journals/twc/JiangGJZZ23}
Y.~Jiang, F.~Gao, M.~Jian, S.~Zhang, and W.~Zhang, ``Reconfigurable intelligent surface for near field communications: Beamforming and sensing,'' \emph{{IEEE} Trans. Wirel. Commun.}, vol.~22, no.~5, pp. 3447--3459, May. 2023.

\bibitem{DBLP:journals/tsp/PizzoTSM22}
A.~Pizzo, A.~de~Jesus~Torres, L.~Sanguinetti, and T.~L. Marzetta, ``Nyquist sampling and degrees of freedom of electromagnetic fields,'' \emph{{IEEE} Trans. Signal Process.}, vol.~70, pp. 3935--3947, Jun. 2022.

\bibitem{DBLP:journals/twc/PizzoSM22}
A.~Pizzo, L.~Sanguinetti, and T.~L. Marzetta, ``Fourier plane-wave series expansion for holographic {MIMO} communications,'' \emph{{IEEE} Trans. Wirel. Commun.}, vol.~21, no.~9, pp. 6890--6905, Jun. 2022.

\bibitem{DBLP:journals/wc/CarvalhoAAAH20}
E.~de~Carvalho, A.~Ali, A.~Amiri, M.~Angjelichinoski, and R.~W.~H. Jr., ``Non-stationarities in extra-large-scale massive {MIMO},'' \emph{{IEEE} Wirel. Commun.}, vol.~27, no.~4, pp. 74--80, Aug. 2020.

\bibitem{DBLP:journals/wcl/ZhangWY22}
Y.~Zhang, X.~Wu, and C.~You, ``Fast near-field beam training for extremely large-scale array,'' \emph{{IEEE} Wirel. Commun. Lett.}, vol.~11, no.~12, pp. 2625--2629, Dec. 2022.

\bibitem{DBLP:journals/icl/GanHYZZ23}
X.~Gan, C.~Huang, Z.~Yang, C.~Zhong, and Z.~Zhang, ``Near-field localization for holographic {RIS} assisted mmwave systems,'' \emph{{IEEE} Commun. Lett.}, vol.~27, no.~1, pp. 140--144, Jan. 2023.

\bibitem{DBLP:journals/ojcs/PizzoSM22}
A.~Pizzo, L.~Sanguinetti, and T.~L. Marzetta, ``Spatial characterization of electromagnetic random channels,'' \emph{{IEEE} Open J. Commun. Soc.}, vol.~3, pp. 847--866, Jun. 2022.

\bibitem{DBLP:journals/dsp/LiuMGLZT20}
H.~Liu, H.~Meng, L.~Gan, D.~Li, Y.~Zhou, and T.~Truong, ``Subspace and sparse reconstruction based near-field sources localization in uniform linear array,'' \emph{Digit. Signal Process.}, vol. 106, p. 102824, Nov. 2020.

\bibitem{DBLP:conf/pimrc/Wang0LHCSA22}
F.~Wang, X.~Wang, X.~Li, X.~Hou, L.~Chen, S.~Suyama, and T.~Asai, ``Ring-type codebook design for reconfigurable intelligent surface near-field beamforming,'' in \emph{2022 {IEEE} 33rd Annual International Symposium on Personal, Indoor and Mobile Radio Communications (PIMRC), Kyoto, Japan, September 12-15, 2022}.\hskip 1em plus 0.5em minus 0.4em\relax {IEEE}, Sep. 2022, pp. 391--396.

\bibitem{DBLP:journals/tcom/LuD23}
Y.~Lu and L.~Dai, ``Near-field channel estimation in mixed los/nlos environments for extremely large-scale {MIMO} systems,'' \emph{{IEEE} Trans. Commun.}, vol.~71, no.~6, pp. 3694--3707, Jun. 2023.

\bibitem{DBLP:journals/cm/ZhangSGDE23}
H.~Zhang, N.~Shlezinger, F.~Guidi, D.~Dardari, and Y.~C. Eldar, ``6g wireless communications: From far-field beam steering to near-field beam focusing,'' \emph{{IEEE} Commun. Mag.}, vol.~61, no.~4, pp. 72--77, Apr. 2023.

\bibitem{7912361}
P.~Nepa and A.~Buffi, ``Near-field-focused microwave antennas: Near-field shaping and implementation,'' \emph{IEEE Antennas and Propagation Magazine}, Jun. 2017.

\bibitem{DBLP:journals/twc/CuiDWZG23}
M.~Cui, L.~Dai, Z.~Wang, S.~Zhou, and N.~Ge, ``Near-field rainbow: Wideband beam training for {XL-MIMO},'' \emph{{IEEE} Trans. Wirel. Commun.}, vol.~22, no.~6, pp. 3899--3912, Jun. 2023.

\bibitem{DBLP:conf/vtc/LiWHCS22}
X.~Li, X.~Wang, X.~Hou, L.~Chen, and S.~Suyama, ``Two-step beamforming scheme for large-dimension reconfigurable intelligent surface,'' in \emph{95th {IEEE} Vehicular Technology Conference, {VTC} Spring 2022, Helsinki, Finland, June 19-22, 2022}.\hskip 1em plus 0.5em minus 0.4em\relax {IEEE}, Jun. 2022, pp. 1--5.

\end{thebibliography}
\end{document}